\documentclass[aps,twocolumn,showpacs,amsmath,amssymb]{revtex4}
\usepackage{graphicx}
\usepackage{dcolumn}
\usepackage{bm}

\begin{document}

\title{Variational procedure for nuclear shell-model calculations 
  and energy-variance extrapolation } 
\author{Noritaka Shimizu$^1$}
\author{Yutaka Utsuno$^2$}
\author{Takahiro Mizusaki$^3$}
\author{Michio Honma$^4$}
\author{Yusuke Tsunoda$^{5}$}
\author{Takaharu Otsuka$^{1,5,6}$}

\affiliation{$^1$ Center for Nuclear Study, University of Tokyo, 
  Hongo Tokyo 113-0033, Japan }
\affiliation{$^2$ Advanced Science Research Center, Japan Atomic Energy Agency, 
  Tokai, Ibaraki 319-1195, Japan }
\affiliation{$^3$ Institute of Natural Sciences, Senshu University, Tokyo, 
  101-8425, Japan}
\affiliation{$^4$ Center for Mathematical Sciences, University of Aizu, Ikki-machi,
  Aizu-Wakamatsu, Fukushima 965-8580, Japan}
\affiliation{$^5$ Department of Physics, University of Tokyo, Hongo,
  Tokyo 113-0033, Japan}
\affiliation{$^6$ National Superconducting Cyclotron Laboratory,
  Michigan State University, East Lansing, Michigan, USA}

\date{\today}

\begin{abstract}
We discuss a variational calculation for nuclear shell-model calculations 
and propose a new procedure for 
the energy-variance extrapolation (EVE) method 
using a sequence of the approximated wave functions obtained by the variational calculation.
The wave functions are described as linear combinations 
of the parity, angular-momentum projected Slater determinants, 
the energy of which is minimized by the conjugate gradient method obeying 
the variational principle. 
The EVE generally works well using the wave functions, 
but we found some difficult cases where the EVE gives 
a poor estimation.
We discuss the origin of the poor estimation 
concerning shape coexistence.
We found that 
the appropriate reordering of the Slater determinants 
allows us to overcome this difficulty and to reduce 
the uncertainty of the extrapolation. 
\end{abstract}


\maketitle

\section{Introduction}

The role of large-scale shell-model calculation has been increasing 
in nuclear structure physics 
with the recent development of faster parallel-computing capabilities.
The most popular method to perform shell-model calculations 
is the Lanczos method and its variants, which can handle $O(10^{11})$ 
configurations.
However, the feasibility of this method is 
still hampered by the exponential increase 
of the Hilbert space of the model space 
as a function of the number of nucleons. 
To overcome this difficulty, 
much effort has been given 
to developing approximation schemes to an exact shell-model 
diagonalization method 
\cite{ppnp_schmid, mcsm_1995, ppnp_mcsm, fda, hybrid_h2, horoi_pci, 
  pittel_dmrg,  roth_importance}. 
These approximation methods truncate the whole Hilbert space to 
a relatively small subspace which is determined 
by various sophisticated methods. 
Therefore, the approximated energy in the subspace 
always provides us with an upper limit to the exact energy
because of the variational principle, 
and an unavoidable small gap remains between the approximated energy 
and the exact energy.

This gap can be removed by extrapolation, 
which has been intensively studied from various perspectives
\cite{horoi_conv, papenbrock_factorize, shen_zhao, 
  extrap_eqs, extrap_2ndorder, extrap_slater, ncsm_extrap, 
  mcsm_extrap, mizusaki_vmcsm}.
The fundamental ingredient of these studies 
is the extrapolation of the approximated eigenenergy
by expanding the subspace into full Hilbert space. 
One realization of this spirit is the Exponential Convergence 
Method (ECM) \cite{horoi_conv}, in which the approximated 
eigenenergy is extrapolated as an exponential function 
of the dimension of the truncated subspace. 
Another scheme of the extrapolation is utilizing the energy variance 
of a sequence of the approximated wave functions 
\cite{extrap_2ndorder}. 
In this scheme, the approximated eigenenergy is 
extrapolated as a polynomial function of 
the corresponding energy variance because 
the energy variance of the exact wave function vanishes. 
The uncertainty of the extrapolated energy is expected to 
be small because the function used for extrapolation 
is a first- or second-order polynomial, 
while the corresponding function is exponential in the ECM. 
This energy-variance extrapolation (EVE) was introduced 
in condensed matter physics \cite{imada_pirg} 
and applied to various approximations in nuclear shell-model calculations 
\cite{extrap_2ndorder, extrap_slater, 
  ncsm_extrap, mcsm_extrap, mizusaki_vmcsm}.

In this article, we report how the EVE method
works with the approximated wave function 
represented by the linear combination of the parity, angular-momentum
projected Slater determinants, such as the Monte Carlo Shell Model (MCSM)
\cite{mcsm_extrap}.
A case in which the EVE is difficult 
is reported in Ref. \cite{okinawa_shimizu}. 
We discuss the origin of this case with relation to shape coexistence 
and how the case is solved by choosing an appropriate reordering of 
the Slater determinants.

We explain the variational calculation to generate
a sequence of approximated wave functions using 
the MCSM and 
conjugate gradient (CG) method  \cite{num_recipe} 
in Sect.\ref{sec:wf}.
In Sect.\ref{sec:simple_eve}, we discuss
how the simple EVE method works.
We also show a typical case in which the EVE shows a poor result.
In Sect.\ref{sec:eve_reord}, we introduce the reordering technique of 
the extrapolation method to remedy this difficulty, 
and demonstrate the feasibility of the method.

\section{Approximated wave functions}
\label{sec:wf}

We briefly describe how to construct the truncated subspace 
to the full Hilbert space in the nuclear shell-model calculations. 
The approximated wave function is written as a linear combination of 
angular-momentum-projected, parity-projected Slater determinants,
\begin{equation}
  |\Psi_N \rangle = \sum_{n=1}^N \sum_{K=-J}^J f^{N}_{n,K} P^{J \pi}_{MK}
  | \phi_n \rangle . 
  \label{eq:wf}
\end{equation}
where $N$ is the number of the basis states, 
which corresponds to the ``MCSM dimension'' in Ref.\cite{okinawa_tabe}.
$P^{J\pi}_{MK}$ is the angular-momentum, parity projector defined as
\begin{equation}
  P^{J\pi}_{MK} =  \frac{1 + \pi \Pi}{2}\frac{2I+1}{8\pi^2} 
  \int d\Omega \  {D^{I}_{MK}}^*(\Omega)
  e^{i\alpha \hat{J}_z} e^{i\beta \hat{J}_y} e^{i\gamma \hat{J}_z} ,
  \label{eq:proj}
\end{equation}
where $\Omega \equiv (\alpha,\beta,\gamma)$ represents the Euler angles, 
and $D^I_{MK}$ denotes Wigner's $D$-function. $\Pi$ denotes 
the parity transformation.
Each $|\phi_n\rangle$ is a deformed Slater determinant, 
\begin{equation}
  |\phi_n \rangle = \prod_k 
  \left( \sum_l D^{(n)}_{lk} c^\dagger_l \right) | - \rangle , 
  \label{eq:slater}
\end{equation}
where $c^\dagger_i$ denotes a creation operator of single particle state $i$
and $|-\rangle$ denotes an inert core.
The coefficient $f^{N}_{n,K}$ is determined 
by the diagonalization of the Hamiltonian matrix in the subspace spanned 
by the projected Slater determinants, $P^{J\pi}_{MK} | \phi_n \rangle $.
This diagonalization also determines the energy, 
$E_N \equiv \langle \Psi_N | H |\Psi_N \rangle$, as a function of $N$.
Note that the dimension of the subspace is the product of the number of states, $N$, 
and the degree of freedom of the $z$-component of angular momentum, $2J+1$.
We increase $N$ until $E_N$ converges enough, 
or the extrapolated energy converges. 

The coefficient $D^{(n)}$ is given by a variational calculation 
utilizing the auxiliary field Monte Carlo technique
and the CG method. 
Each $D^{(n)}$ is 
determined by minimizing $E_{N=n}$ while keeping 
$D^{(1)}, D^{(2)}, ... D^{(n-1)}$, 
in a manner similar to the MCSM \cite{ppnp_mcsm}, 
the few-dimensional basis approximation \cite{fda}, 
and the Hybrid Multideterminant method \cite{hybrid_h2}. 
We perform the MCSM procedure in a small number of steps 
to get the initial states of the CG process in order to avoid 
a trap by local minima.
Since the $D^{(n)}$ is determined sequentially, hereafter, 
we call this procedure 
the Sequential Conjugate Gradient (SCG) method.

The energy variance of the approximated wave function $|\Psi_N\rangle$ 
is also evaluated as 
$\langle \Delta H^2 \rangle_N = \langle \Psi_N | H^2 |\Psi_N \rangle - E_N^2 $.
The energy, energy gradient, and energy variance are evaluated
under the angular-momentum, parity projection technique throughout this work. 
In this sense, our method is a ``variation after projection''.

\section{Anomalous kink in the EV plot}
\label{sec:simple_eve}

\begin{figure}[htbp]
  \includegraphics[scale=0.4]{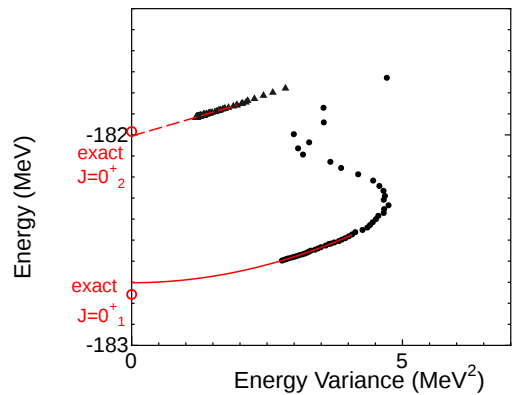}
  \caption{(Color online) Energy vs. energy variance plot 
    of the ground state and $0^+_2$ state 
    of $^{72}$Ge by the SCG method. 
    The circles and triangles 
    show the energies and energy variances of $0^+_1$ and 
    $0^+_2$, respectively. The lines are fitted for these points 
    using a second-order polynomial. 
  }
  \label{ge72_wo_rd}
\end{figure}

The $^{72}$Ge is a typical case such that the relation 
between the energy 
and its variance is not monotonic, 
which is reported in Ref.\cite{okinawa_shimizu}. 
This nucleus exhibits a feature of shape coexistence 
due to the $N=40$ magicity \cite{jun45},
and this is considered to be a main reason for ill-behavior 
as will be discussed later.
In the $^{72}$Ge shell-model calculation, 
we take the $f_5pg_9$-shell, which consists of $0f_{5/2}$, $1p_{3/2}$,
$1p_{1/2}$, and $0g_{9/2}$ single-particle orbits, as a model space 
and the effective interaction JUN45 is used \cite{jun45}. 
The $m$-scheme dimension of $^{72}$Ge reaches 140,050,484, 
which can be handled directly by recent shell-model diagonalization codes.
The exact values in Fig.  \ref{ge72_wo_rd} represent the results 
of the shell-model diagonalization utilizing 
the code MSHELL64 \cite{mshell64}.

Figure \ref{ge72_wo_rd} shows the energy and 
energy variance of $|\Psi_N\rangle$ with $1\leq N\leq 100$ for 
the $0^+_1$ state. 
The $|\Psi_N\rangle$ is obtained by the SCG method. 
Hereafter, we call the plot of the energy as a function of a variance 
``EV plot''.
In the EV plot concerning the $0^+_1$ state, 
one can find a similar anomalous kink 
in the case of the MCSM  \cite{okinawa_shimizu}. 
This kink reduces the region available for the second order fit, 
and deteriorates the certainty of the extrapolation, 
resulting in a $50$ keV overestimation.
To obtain the $0^+_2$ energy, additional 100 bases are generated 
by the SCG method for minimizing $E_N(0^+_2)$.
The dashed line in Fig. \ref{ge72_wo_rd} is 
fitted for the points concerning $J=0^+_2$,
whose extrapolation agrees with the exact energy well, 
but a small underestimation remains.

\begin{figure}[htbp]
  \includegraphics[scale=0.4]{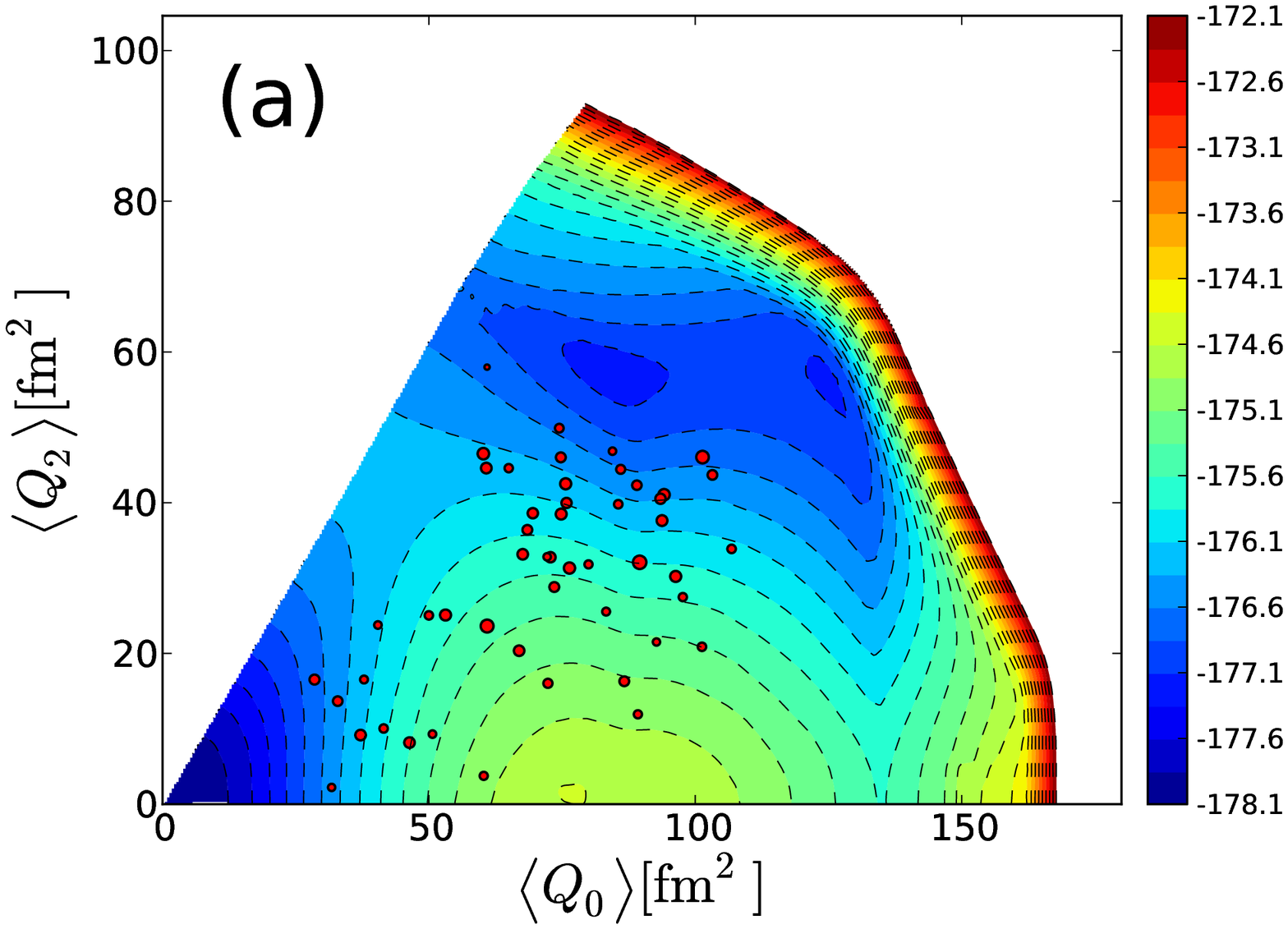}
  \includegraphics[scale=0.4]{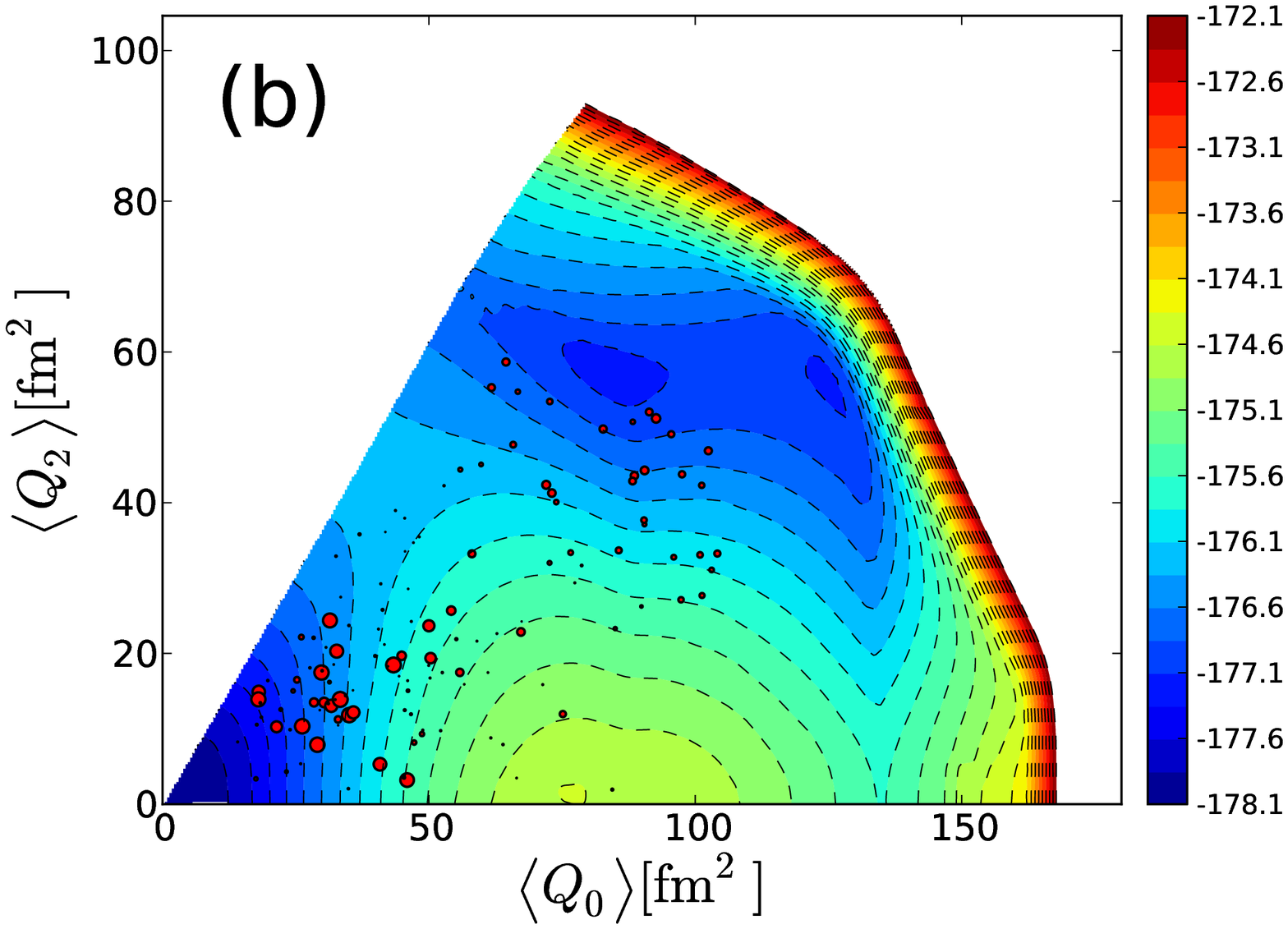}
  \caption{(Color online) Total energy surface of the $^{72}$Ge by 
    the $Q$-constrained Hartree-Fock calculation.
    The scattered circles denote the deformation of 
    the basis states of the SCG wave functions 
    for (a) $0^+$ ground state, and (b) second $0^+$ state. 
    The contour width is 250 keV.}
  \label{ge72_pes}
\end{figure}

Figure \ref{ge72_pes} shows the total energy surface 
provided by the $Q$-constrained Hartree-Fock calculation 
\cite{ringschuck} using the same shell-model Hamiltonian. 
There are two low-energy regions 
corresponding to the shape coexistence phenomenon \cite{jun45}.
In order to discuss the intrinsic structures of the $0^+_1$ and 
$0^+_2$ wave functions, 
we plot the deformation of each unprojected basis state $|\phi_n\rangle$ 
of the SCG wave function $|\Psi_N\rangle$ in Fig. \ref{ge72_pes}.
The location of the scattered circles shows the 
quadrupole deformation, namely, $\langle \phi_n |Q_0| \phi_n\rangle $
and  $\langle \phi_n |Q_2 |\phi_n \rangle$
where $Q_M$ is an $M$-component of the mass quadrupole operator 
and $|\phi_n\rangle$ is rotated so that
$\langle \phi_n |Q_{\pm 1} |\phi_n \rangle = 0$. 
It is interesting to see that the circles scatter 
in a broad region of the 
energy surface, not near the local minima, 
but on the hillside and triaxially deformed regions
due to the effect of the configuration mixing 
and the ``variation after projection''.

The area of each scattered circle on Figs. \ref{ge72_pes} (a) and (b) 
is proportional to the overlap probability of each projected basis state
with the resulting many basis states, 
$\frac{1}{\cal N} |\langle \Psi_N| P^{J}|\phi_n\rangle |^2$, 
where $\cal N$ is the normalization factor, 
${\cal N}  =  \langle \phi_n| P^{J}|\phi_n\rangle $.
Concerning the $0^+_1$ state shown in Fig. \ref{ge72_pes} (a), 
the overlap probability is rather small and 0.54 at most. 
The circles with relatively large overlap scatters 
in a broad region of the energy surface.
The overlap probability with $P^J|\phi_1\rangle$, e.g. the Hartree-Fock solution 
with the variation after projection is 0.32, which is modest.
This overlap implies that the $0^+_1$ state is 
described by a linear combination of 
a relatively large number of basis states essentially, 
or the effect of the configuration mixing plays an important role.
On the other hand, in Fig. \ref{ge72_pes} (b), 
the points of the large overlap 
concentrate near the spherical region concerning the $0^+_2$ state. 
The overlap probabilities corresponding to these points
are large with the $0^+_2$ many-basis state, 
and the probability is 0.67 at most. 
This means that the wave function can be well approximated by 
a few number of projected Slater determinants, 
or mean-field description works 
far better than in the case of the  $0^+_1$ state. 
This property makes the first few SCG basis states 
for minimizing $0^+_1$ energy 
dominated by the  $0^+_2$ state, not by the $0^+_1$ state. 
These different properties of the $0^+_1$ and $0^+_2$ wave functions 
give rise to the different behavior of energy convergence 
as a function of the number of basis states,
which makes extrapolation with a number of basis states 
difficult \cite{okinawa_shimizu}.  

In Fig. \ref{ge72_wo_rd}, 
the variance shows the local minimum at $E=-182$ MeV, 
which is near the energy of the $0^+_2$ state. 
This figure indicates that the wave function 
comprised of the first 10 bases are 
dominated by the true $0^+_2$ state, not the $0^+_1$ state. 
This is consistent with the discussion using the energy surface 
in the previous paragraph.

Such an anomalous kink in the EV plot
is also seen in the $8^+_1$ state of $^{56}$Ni 
with the FPD6 effective interaction \cite{fpd6}. 
$^{56}$Ni  is known to have shape coexistence \cite{ni_coex_mcsm}, 
which is consistent with the previous discussion.
This situation may occur in the case where
the next lowest energy eigenvalue 
is close to the target one and the mean-field solution favors the 
next lowest state.

\section{EVE and reordering of basis states}
\label{sec:eve_reord}

The emergence of the anomalous kink 
discussed in Sect.\ref{sec:simple_eve} 
gives rise to a large uncertainty of the EVE. 
In order to remove such a kink and to improve the precision, 
we introduce the reordering of basis states of the SCG wave function, 
represented in Eq.(\ref{eq:wf}). 

The SCG wave function comprises a set of $N_m$ basis states, 
$|\phi_n\rangle$ with $1\le n \le N_m$.
These basis states also give us 
a sequence of approximated wave functions, 
$|\Psi_N\rangle$ with $1\le N \le N_m$, 
which is used for the simple EVE method. 
Meanwhile, reordering these $N_m$ basis states
using a permutation, $\sigma(n)$, yields 
another sequence of approximated wave functions, 
\begin{equation}
  |\Psi_N^{\rm (ro)} \rangle = \sum_{n=1}^N \sum_{K=-J}^J 
  f^{N{\rm (ro)}}_{n,K} P^{J \pi}_{MK}
  | \phi^{\rm (ro)}_n \rangle
  \label{eq:reorder-wf}
\end{equation}
where  $|\phi^{\rm (ro)}_n \rangle = | \phi_{\sigma(n)} \rangle$.
The corresponding energy $E^{\rm (ro)}_N$, 
and the energy variance $\langle \Delta H^2\rangle^{\rm (ro)}_N$
provide us with a new EV plot. 
By reordering the basis states with $\sigma(n)$, 
we can simulate another truncation scheme.
Because the behavior of the fitted line in the EV plot 
depends on the truncation scheme \cite{extrap_eqs}, 
we remove an anomalous behavior of the EV plot 
and make the extrapolation stable by using an appropriate 
$\sigma(n)$.
We describe the way how to obtain an appropriate order of 
basis states in this section.

\subsection{Procedure of reordering technique}
\label{sec:reordering}

The relation of the energy and its variance is usually assumed 
to be expressed as a second-order polynomial. 
Because a second-order-term error is roughly estimated 
as $\delta c_2 (\langle \Delta H^2 \rangle_N)^2$, 
where $\delta c_2$ is a second-order-term error of $\chi$-square fitting, 
the uncertainty of the second order term 
often causes a relatively large error of the extrapolated value 
in case $\langle \Delta H^2 \rangle_N$, is not small enough. 
If we can find an order of the basis states in which the fitted curve 
is a first-order polynomial, the error of the extrapolation 
will come mainly from the coefficient of the first-order term, 
which can be estimated as roughly proportional to 
$\delta c_1 \langle \Delta H \rangle_N$, 
where $\delta c_1$ is the first-order-term error of $\chi$-square fitting. 
This error should be far smaller than that of second-order polynomial
if $\langle \Delta H^2 \rangle_N$ is large.
Thus, our strategy is to select the order of 
the basis states, in which the fitted curve is close to linear, 
by changing the order of the basis states.

Now, we discuss the relation between energy difference and energy variance
following the idea of Ref.\cite{extrap_eqs}. 
We define the energy difference $\delta E$ between the energy 
expectation value $E = \langle \Psi| H | \Psi \rangle $ 
of the approximated wave function $|\Psi\rangle$ 
and the lowest exact energy eigenvalue $E_0$ as 
\begin{equation}
  \delta E = \langle \Psi |H|\Psi \rangle - E_0 , 
\end{equation}
and the energy variance of the approximated wave function as 
\begin{equation}
  \langle \Delta H^2 \rangle = \langle \Psi |H^2|\Psi \rangle 
  - \langle \Psi |H|\Psi \rangle^2 .
\end{equation}

An approximate ground state $|\Psi \rangle$ can be decomposed into 
the exact eigenstate, $|\psi_0\rangle$, 
and the rest of the component, $|\psi_r\rangle$ as 
\begin{equation}
  | \Psi \rangle = c |\psi_0\rangle + d |\psi_r \rangle, 
\end{equation}
where $c^2 + d^2 = 1$. $|\psi_r\rangle$ is 
expanded by the exact excited states such as
\begin{equation}
  | \psi_r \rangle = \sum_{n\neq 0} c_n | \psi_n\rangle . 
\end{equation}
By defining the moments $D_j$ as 
\begin{equation}
  D_j = \sum_{n\neq 0} c_n^2 (E_n-E_0)^j , 
\end{equation}
we obtain 
\begin{equation}
  \delta E = d^2 D_1 , 
\end{equation}
\begin{equation}
  \langle \Delta H^2 \rangle  = d^2 D_2 - (d^2D_1)^2 .
  \label{eq:poly2h2}
\end{equation}
By eliminating $d^2$, we obtain
\begin{equation}
  \langle \Delta H^2 \rangle  =  \frac{D_2}{D_1} \delta E - (\delta E)^2 .
\end{equation}
In Eq. (\ref{eq:poly2h2}), 
$\langle \Delta H^2 \rangle$ is written as a second-order 
polynomial of $ \delta E$, 
which explains the parabola shape of the EV plot 
in Fig. \ref{ge72_wo_rd} (circles below -182MeV).
The authors of Ref.\cite{extrap_eqs} solve $\delta E$ as a function of 
$\langle \Delta H^2 \rangle$, which is shown to be approximated 
as a second-order polynomial.
Now, we assume $\delta E$ is proportional to $\langle \Delta H^2 \rangle$, 
which is realized in case $\frac{D_2}{D_1} \rightarrow \infty$. 
Because of 
$\frac{\partial (\delta E)}{\partial  \langle \Delta H^2 \rangle}|_{\delta E=0} = \frac{D_1}{D_2}$, 
we reorder the basis states so that 
$\frac{\partial (\delta E)}{\partial \langle \Delta H^2 \rangle}$ 
is as small as possible, resulting in linear proportionality. 

In practice, we perform the following procedure:
\begin{enumerate}
\item A fixed number, $N_m$, of the basis states is 
  obtained by the SCG method. 
\item Choose the $N_m$-th basis state $|\phi^{\rm (ro)}_{N_m}\rangle$  
  from $N_m$ candidates by minimizing 
  $\frac{E^{\rm (ro)}_{N_m-1} - E^{\rm (ro)}_{N_m}}
  {\langle \Delta H^2 \rangle^{\rm (ro)}_{N_m-1} 
    - \langle \Delta H^2 \rangle^{\rm (ro)}_{N_m}}$ .
\item Set $N$ as $N_m-1$. 
  Choose the $N$-th basis state, $|\phi^{\rm (ro)}_N\rangle$ 
  from $N$ candidates 
  (basis states except for already fixed states, 
  $|\phi^{\rm (ro)}_{N'}\rangle$ with $N+1 \leq N' \leq N_m$   )
  by minimizing 
  $\frac{E^{\rm (ro)}_{N-1} - E^{\rm (ro)}_{N}}
  {\langle \Delta H^2 \rangle^{\rm (ro)}_{N-1} 
    - \langle \Delta H^2 \rangle^{\rm (ro)}_{N}}$ .
\item Set $N$ as $N-1$. The previous step is iterated recursively up to 
  determining the first state, $|\phi^{\rm (ro)}_1\rangle$.
\end{enumerate}

Note that this procedure needs no additional heavy computation 
to the SCG, 
because the matrix elements of 
energy and its variances are evaluated once and stored, 
and we only require the diagonalization of the matrix whose dimension is 
$N \le N_m \simeq 100$ 
for each candidate order of the basis states.

As a result, we obtain a fitted line which is closest to linear 
in the provided set of basis states. 
Because the reordering makes the gradient of 
the EV plot as small as possible, 
the anomalous kink discussed in Sect. \ref{sec:simple_eve} vanishes, 
which is demonstrated in the following subsection.

\subsection{$^{72}$Ge in $f_5pg_9$-shell}
\label{sec:eve_ge72}

\begin{figure}[htbp]
  \includegraphics[scale=0.4]{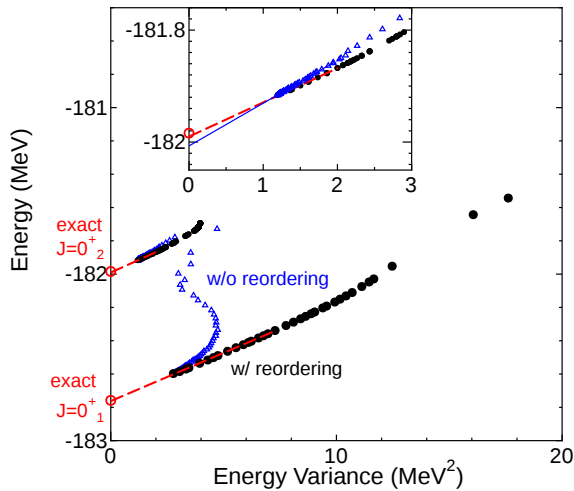}
  \caption{(Color online) Energy vs. energy variance plot 
    for the ground state and $0^+_2$ state 
    of $^{72}$Ge in $f_5pg_9$-shell. 
    The open triangles 
    are obtained by the CG method 
    without reordering 
    and the solid circles are with reordering.
    The inset shows a magnified view 
    for the $0^+_2$ energy around $\langle \Delta H^2\rangle\simeq 0$.
    See text for detail.
  }
  \label{ge72_cg}
\end{figure}

We apply the reordering technique to the $^{72}$Ge in $f_5pg_9$-shell. 
The filled circles in Fig. \ref{ge72_cg} shows the EV plot 
with the reordering technique. 
The anomalous kink of the $0^+_1$ state vanishes
in the plot with reordering, 
and the point moves smoothly and approaches the exact energy on the $y$-axis 
as usual \cite{mcsm_extrap}.
These points are fitted by a first-order polynomial, which is shown 
as a dashed line. 
Unlike the fit without reordering, 
these points are on the line in a large range of the energy variance
which can be used for the fitting. 
This makes the extrapolation procedure stable.

\begin{figure}[htbp]
  \includegraphics[scale=0.4]{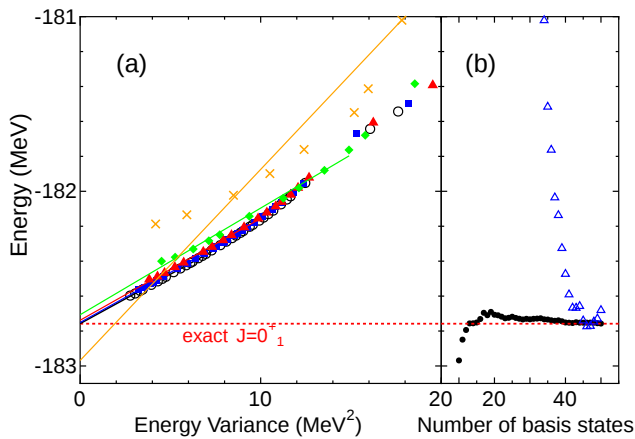}
  \caption{(Color online) (a) Energy vs. energy variance plot of the ground state energy 
    of $^{72}$Ge in the $f_5pg_9$-shell with the SCG wave function 
    and reordering. 
    $N_m=10$ (orange crosses), $20$ (green diamonds),
    $30$ (red triangles), $40$  (blue squares),  and $50$ (open black circles). 
    (b) Extrapolated value vs. the number of basis states, $N_m$. 
    The exact energy is also shown by the dotted line. 
    The solid points and open blue triangles denote 
    the extrapolated energies with first-order fitting and reordering, 
    and those with second-order fitting without reordering, respectively.
  }
  \label{ge72_seq_b1050}
\end{figure}

Figure \ref{ge72_seq_b1050}(a) shows the EV plot provided by 
the SCG method with a various number of basis states, 
$N_m=10, 20, 30, 40$ and $50$. 
At each $N_m$, the last 10 points are used to make the fitted line.
While the extrapolated values of $N_m=10$ apparently underestimate 
the ground state energy, the value converges as a function of $N_m$ 
for $N_m\geq 30$. 

Discussing the stability of 
the extrapolated value is difficult 
because all the energy-variance points 
with reordering change by increasing the $N$. 
To discuss this stability, we show the extrapolated 
energy itself with the reordering technique as a function of $N_m$.
Figure \ref{ge72_seq_b1050} (b) shows the convergence property of 
the extrapolated energy as a function of $N_m$. 
The extrapolated energy with reordering using the last 10 points, 
or $(N_m-9)$-th, $(N_m-8)$-th, ... $N_m$-th points, 
converges well 
within a few keV for  $N_m>40$.
On the other hand, the extrapolated value without reordering 
shows a large fluctuation due to the kink in the EV plot. 

The inset of Fig. \ref{ge72_cg} shows the EV plot 
concerning $J=0^+_2$. 
The open triangles and the fitted line 
show the energy and variance without reordering. 
Its extrapolation underestimates 20 keV, the exact value, 
while the extrapolated value with the reordering technique
agrees with the exact one within the 6 keV error.
The reordering technique reduces the error of the extrapolation 
even in the case of the simple EVE without reordering, 
and works well.

\subsection{$^{64}$Ge in $pfg9$-shell}
\label{sec:pfg9}

\begin{figure}[htbp]
  \includegraphics[scale=0.4]{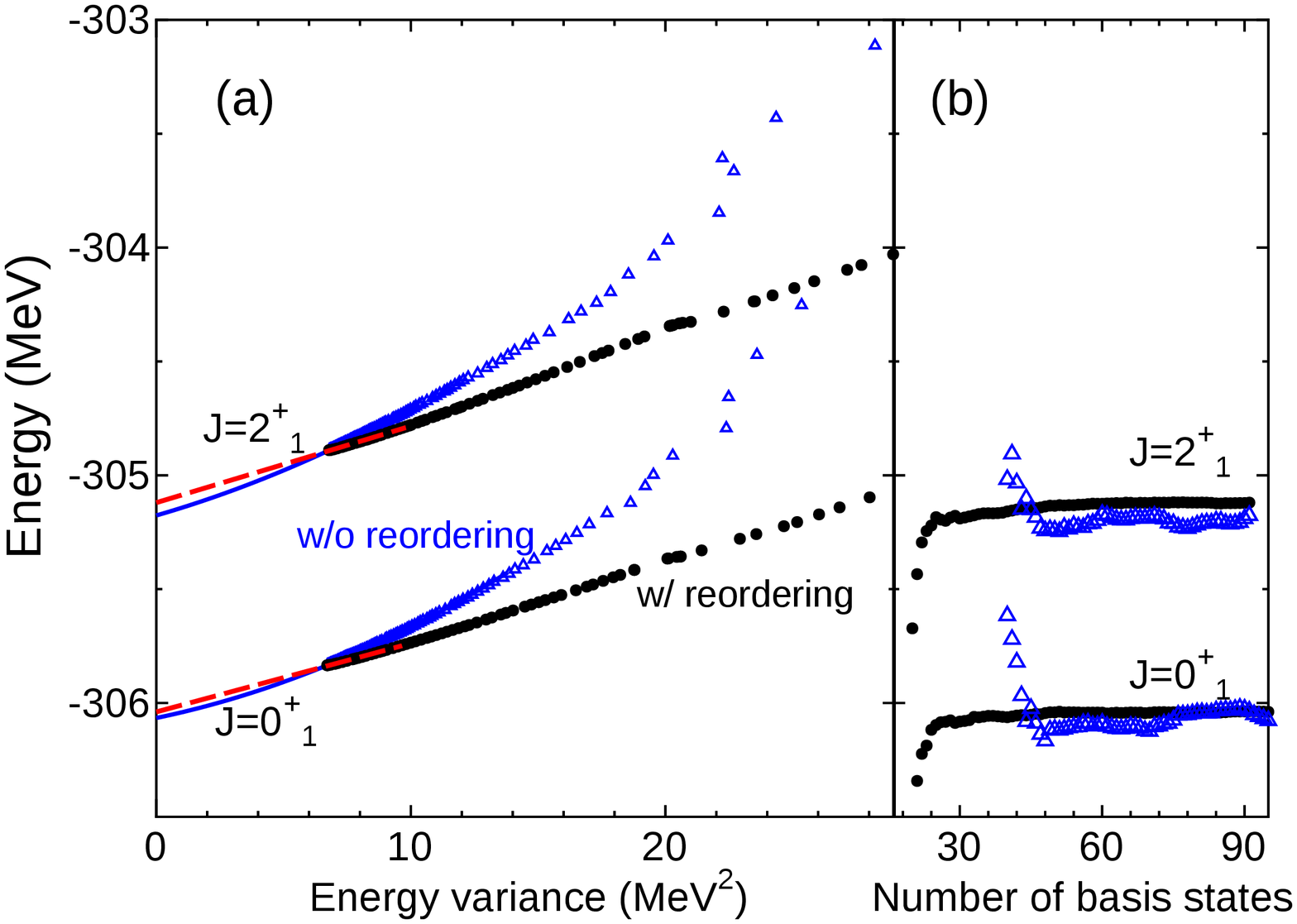}
  \caption{(Color online) (a) Energy vs. energy variance plot 
    for the $0^+_1$ and $2^+_1$ states of $^{64}$Ge 
    in the $pfg9$-shell 
    obtained by the SCG method. 
    (b) Extrapolated energies with reordering (filled circles) 
    and without reordering (open triangles) 
    as functions of $N_m$. }
  \label{fig:ge64_cg}
\end{figure}

In Ref. \cite{mcsm_extrap}, we show a demonstration 
of the validity of the MCSM and energy-variance by 
$^{64}$Ge in the $pf+g_{9/2}$-shell model space, which 
consists of $0f_{7/2}, 0f_{5/2}, 1p_{3/2}, 1p_{1/2}$ and $0g_{9/2}$ 
single-particle orbits. 
We use the PFG9B3 effective interaction \cite{pfg9b3}, 
which was also used in Refs. \cite{horoi_leveldens, mcsm_extrap}.
The $m$-scheme dimension of the system reaches $1.7\times 10^{14}$, 
which cannot be reached by the conventional Lanczos diagonalization technique. 
In the present work, this nucleus is taken as an example again. 
We show the results using 
the SCG method and by extrapolation with the reordering technique
in Fig. \ref{fig:ge64_cg}. 
As you can see, the extrapolated energies with reordering technique 
well converge in a relatively small number of basis states. 
In this case, the energies with reordering agree 
well with those without reordering, and those of the MCSM method 
which are shown in Ref.\cite{mcsm_extrap}.

\section{Summary}

We have discussed variational procedures, the SCG method, 
for nuclear shell-model calculations. 
Based on the SCG wave functions, we propose a new procedure 
of energy variance extrapolation, which can solve a complex 
problem due to shape coexistence.

For example, the simple EVE is difficult 
to solve at the $0^+_1$ state of $^{72}$Ge with the JUN45 interaction.
In this case, the anomalous kink appears in the EV plot, which
shows the transition of the approximated wave function 
from the second lowest state to the lowest state. 
We discussed its origin in view of shape coexistence,  
and found that the reordering technique of the basis states 
allows us to extrapolate the eigenenergy successfully even in this case.

We demonstrated that the reordering technique in the EV plot 
allows us to make a linear fit; 
and therefore, 
the reordering of the basis states makes the extrapolation procedure stable 
and suppresses the uncertainties of the extrapolated value. 
This procedure is expected to be quite useful 
into performing the precise estimate of nuclear energies 
based on large-scale shell-model calculations
and no-core shell-model calculations \cite{okinawa_tabe}.
For the estimatation of observables other than energy, 
such as quadrupole moment, 
the order which was used in the present work
is not suitable for the extrapolation procedure. 
We are investigating a way to determine 
an order suitable for the extrapolation 
of these observables.

\begin{acknowledgments}
We acknowledge Dr. T. Abe for fruitful discussions. 
This work has been supported by a Grant-in-Aid 
for Young Scientists (20740127) from JSPS, 
the SPIRE Field 5 from MEXT, and the CNS-RIKEN joint project 
for large-scale nuclear structure calculations. 
The numerical calculation was performed mainly 
on the T2K Open Supercomputers at the University of Tokyo 
and Tsukuba University. 
The exact shell-model calculation was performed 
by the code MSHELL64 \cite{mshell64}.

\end{acknowledgments}



\begin{thebibliography}{99}

\bibitem{mcsm_1995} M. Honma, T. Mizusaki, and T. Otsuka, 
  Phys. Rev. Lett. {\bf 75}, 1284 (1995). 

\bibitem{ppnp_mcsm}  T. Otsuka, M. Honma, T. Mizusaki, N. Shimizu, 
  and Y. Utsuno, Prog. Part. Nucl. Phys. {\bf 47}, 319 (2001).

\bibitem{fda} M. Honma, B. A. Brown, T. Mizusaki, and T. Otsuka, 
  Nucl. Phys. A {\bf 704}, 134c (2002), M. Honma, T. Otsuka, B.A. Brown and 
  T. Mizusaki,  Phys. Rev. C \textbf{65}, 061301(R) (2002).

\bibitem{ppnp_schmid} K. W. Schmid, Prog. Part. Nucl. Phys. 
  {\bf 52}, 565 (2004).

\bibitem{pittel_dmrg} S. Pittel and N. Sandulescu, Phys. Rev. C \textbf{73}, 
  014301 (2006).

\bibitem{hybrid_h2} G. Puddu, J. of Phys. G: Nucl. Part. Phys.  {\bf 32}  321 (2006), 
  G. Puddu, Eur. Phys. J. A {\bf 34}, 413 (2007).

\bibitem{roth_importance} R. Roth, and P. Navratil, Phys. Rev. Lett. {\bf 99} 092501 (2007) 

\bibitem{horoi_pci} Z.-C. Gao, M. Horoi, and Y. S. Chen,   
  Phys. Rev. {\bf C 80}, 034325 (2009)

\bibitem{horoi_conv} M. Horoi, A. Volya, and V. Zelevinsky, 
  Phys. Rev. Lett. {\bf 82}, 2064 (1999); M. Horoi, B. A. Brown, and V. Zelevinsky, 
  Phys. Rev. C \textbf{67} 034303 (2003).

\bibitem{papenbrock_factorize} T. Papenbrock and D. J. Dean, 
  Phys. Rev. C \textbf{67}, 051303(R) (2003).

\bibitem{shen_zhao} J.J. Shen, Y. M. Zhao, A. Arima,
  and N. Yoshinaga, Phys. Rev. C \textbf{83}, 044322 (2011).

\bibitem{extrap_2ndorder} T. Mizusaki and M. Imada, 
  Phys. Rev. C {\bf 65}, 064319 (2002).

\bibitem{extrap_eqs} T. Mizusaki and M. Imada,  Phys. Rev. C 
  {\bf 67}, 041301 (2003). 

\bibitem{mizusaki_vmcsm} T. Mizusaki and N. Shimizu, 
  Phys. Rev. C \textbf{85}, 021301(R) (2012).


\bibitem{extrap_slater} T. Mizusaki, Phys. Rev. {\bf C 70}, 044316 (2004).

\bibitem{ncsm_extrap} H. Zhan, A. Nogga, B.R. Barrett, J.P. Vary 
  and P. Navratil, Phys. Rev. C {\bf 69}, 034302 (2004).

\bibitem{mcsm_extrap} N. Shimizu, Y. Utsuno, T. Mizusaki, T. Otsuka, T. Abe, 
  and M. Honma, Phys. Rev. C \textbf{82}, 061305(R) (2010).

\bibitem{imada_pirg} M. Imada and T. Kashima, 
  J. Phys. Soc. Jpn. {\bf 69} 2723 (2000).

\bibitem{okinawa_shimizu} N. Shimizu, Y. Utsuno, T. Mizusaki, 
  T. Otsuka, T. Abe and M. Honma, 
  AIP Conf. Proc. {\bf 1355}, 138 (2011).

\bibitem{num_recipe} Numerical Recipes in Fortran 77, the Art of Scientific Computing, 
  2nd ed., Cambridge University Press: Cambridge, (1992).

\bibitem{okinawa_tabe} T. Abe, P. Maris, T. Otsuka, N. Shimizu, Y. Utsuno, 
  and J. P. Vary,   AIP Conf. Proc. {\bf 1355}, 138 (2011).

\bibitem{jun45} M. Honma, T. Otsuka, T. Mizusaki, and M. Hjorth-Jensen, 
  Phys. Rev. {\bf C 80}, 064323 (2009).

\bibitem{mshell64} T. Mizusaki, N. Shimizu, Y. Utsuno, and M. Honma,
code MSHELL64, unpublished.

\bibitem{ringschuck} P. Ring and P. Schuck, {\it The Nuclear Many-Body Problem}, 
  Springer-Verlag, New York, 1980.

\bibitem{fpd6} W.A. Richter, M.G. van der Merwe, R.E. Julies and B.A. Brown, 
  Nucl. Phys. {\bf A523}, 325,  (1991).

\bibitem{ni_coex_mcsm} T. Mizusaki, T. Otsuka, Y. Utsuno, M. Honma 
  and T. Sebe, Phys. Rev. C \textbf{59} R1846 (1999).

\bibitem{pfg9b3} M. Honma {\it et al.}, unpublished.

\bibitem{horoi_leveldens}  R. A. Sen'kov and M. Horoi, 
  Phys. Rev. {\bf C 82} 024304 (2010).


\end{thebibliography}
\end{document}